\documentclass[sigconf, final]{acmart}
\usepackage{graphicx}
\usepackage{amsmath}
\usepackage{booktabs}
\usepackage{algorithm}
\usepackage{algorithmic}
\usepackage{amsfonts}
\usepackage{multirow}
\usepackage{makecell}
\usepackage{subfigure}
\usepackage{color}
\usepackage{bm}
\usepackage{epstopdf}
\usepackage{url}
\usepackage[cal=cm]{mathalfa}
\usepackage{balance}
\usepackage{threeparttable}
\usepackage{lipsum}
\usepackage{enumitem}
\usepackage{pifont}
\usepackage{tabularx}
\usepackage{makecell}
\usepackage{float}
\usepackage[table]{xcolor}
\usepackage{array}
\usepackage{afterpage}

\setlist[itemize]{leftmargin=*}

\AtBeginDocument{%
  \providecommand\BibTeX{{%
    \normalfont B\kern-0.5em{\scshape i\kern-0.25em b}\kern-0.8em\TeX}}}

\renewcommand\footnotetextcopyrightpermission[1]{}

\begin{document}
\title{OnePiece: The Great Route to Generative Recommendation\\---A Case Study from Tencent Algorithm Competition}

\author{Jiangxia Cao, Shuo Yang, Zijun Wang, Qinghai Tan}
\affiliation{
  \institution{OnePiece Team, NJUST}
  \country{jiangxiacao@gmail.com}
 }

\renewcommand{\shorttitle}{OnePiece}

\begin{abstract}
% % 
In past years, the OpenAI's Scaling-Laws shows the amazing intelligence with the next-token prediction paradigm in neural language modeling, which pointing out a free-lunch way to enhance the model performance by scaling the model parameters.
In RecSys, the retrieval stage is also follows a 'next-token prediction' paradigm, to recall the hunderds of items from the global item set, thus the generative recommendation usually refers specifically to the retrieval stage (without Tree-based methods).
This raises a philosophical question: without a ground-truth next item, does the generative recommendation also holds a potential scaling law?
In retrospect, the generative recommendation has two different technique paradigms:
(1) ANN-based framework, utilizing the compressed user embedding to retrieve nearest other items in embedding space, e.g, Kuaiformer.
(2) Auto-regressive-based framework, employing the beam search to decode the item from whole space, e.g, OneRec.
In this paper, we devise a unified encoder-decoder framework to validate their scaling-laws at same time.
Our empirical finding is that both of their losses strictly adhere to power-law Scaling Laws ($R^2$>0.9) within our unified architecture. 
Experiments on TencentGR-100M/10M demonstrate the superiority of our method, and we ranked 10th in the preliminary round, 9th in the semi-finals, and 13th in the finals among thousands of scale teams~\footnote{Our codes:\url{https://github.com/shuoyang2/OnePiece}}.
\end{abstract}

\maketitle

\section{Introduction}

\textbf{Background.}
Scaling Laws~\cite{achiam2023gpt} have emerged as a core driver of breakthroughs in domains such as Natural Language Processing (NLP) and Computer Vision (CV). 
However, conventional recommender system paradigms, particularly those based on discriminative models, have struggled to replicate similar success.
A typical observation is that despite substantial efforts from both industry and academia to devise increasingly sophisticated architectures—evolving from early models like DIN~\cite{zhou2018deep} to MIMN~\cite{pi2019practice} and SIM~\cite{pi2020search}—the performance enhancements exhibit a clear trend of diminishing returns. 
This phenomenon of performance saturation strongly suggests that merely increasing the complexity of discriminative models may be approaching a fundamental paradigm ceiling.
In stark contrast to the dilemma faced by discriminative models, generative models, epitomized by Large Language Models (LLMs), have demonstrated remarkable scalability.
It shifts the research focus from the traditional discriminative task of "predicting click-through rates" to exploring the generative task of "next interactive item"\cite{sasrec}.

\textbf{Related work.}
Recommender systems have long been dominated by discriminative models. 
Over the past decade, this paradigm has dominated the landscape, evolving from early Matrix Factorization to deep architectures like DeepFM~\cite{guo2017deepfm}, and subsequently to attention-based models such as DIN~\cite{zhou2018deep} and SIM~\cite{pi2020search} for modeling user behavior sequences.
Inspired by the remarkable scaling effects of LLMs in NLP, pioneering works like Kuaiformer~\cite{liu2024kuaiformer} unify the Transformer as a backbone for industrial-scale multi-interests retrieval. 
Subsequent approaches, such as OneRec~\cite{deng2025onerec}, further reformulate recommendation as an auto-regressive generation process, validating continuous performance growth with larger parameter scales. 
These advancements suggest that generative architectures adhering to Scaling Laws are poised to become the core paradigm for next-generation recommender systems.

\textbf{Motivation.}
While generative recommendation (e.g., OneRec~\cite{deng2025onerec}) has successfully adopted Scaling Laws to capture open-ended user interests, it relies on beam search to decode the next group of items.
However, for the heavy inference cost, the topk in the middle stage has considerable limitations, and the logp path probability will be affected by highly popular items.
% probabilistic sampling (e.g., ) inherently limits both ranking precision and inference speed.
% 
In contrast, widely-used embedding-based retrieval offers efficient and precise logQ debiased similar scores but struggles to explore an item's potential in the total space within a single query head. 
% match the continuous scaling potential of generative backbones due to capacity constraints, e.g., limited by the batch size and model paradigm at the same time.
% 
Therefore, we consider connecting them in a unified framework: first, utilizing the beam-search technique to identify the item group from the whole space and then scoring them unbiased by the embedding-based technique, for a broad and precise retrieval within a limited quota.
% there is a need for a framework that retains the massive capacity of generative backbones while enforcing the precision and efficiency of Embedding-Based retrieval mechanisms suitable for industrial deployment.

\begin{figure}[t!]
\begin{center}
\includegraphics[width=8cm,height=2.5cm]{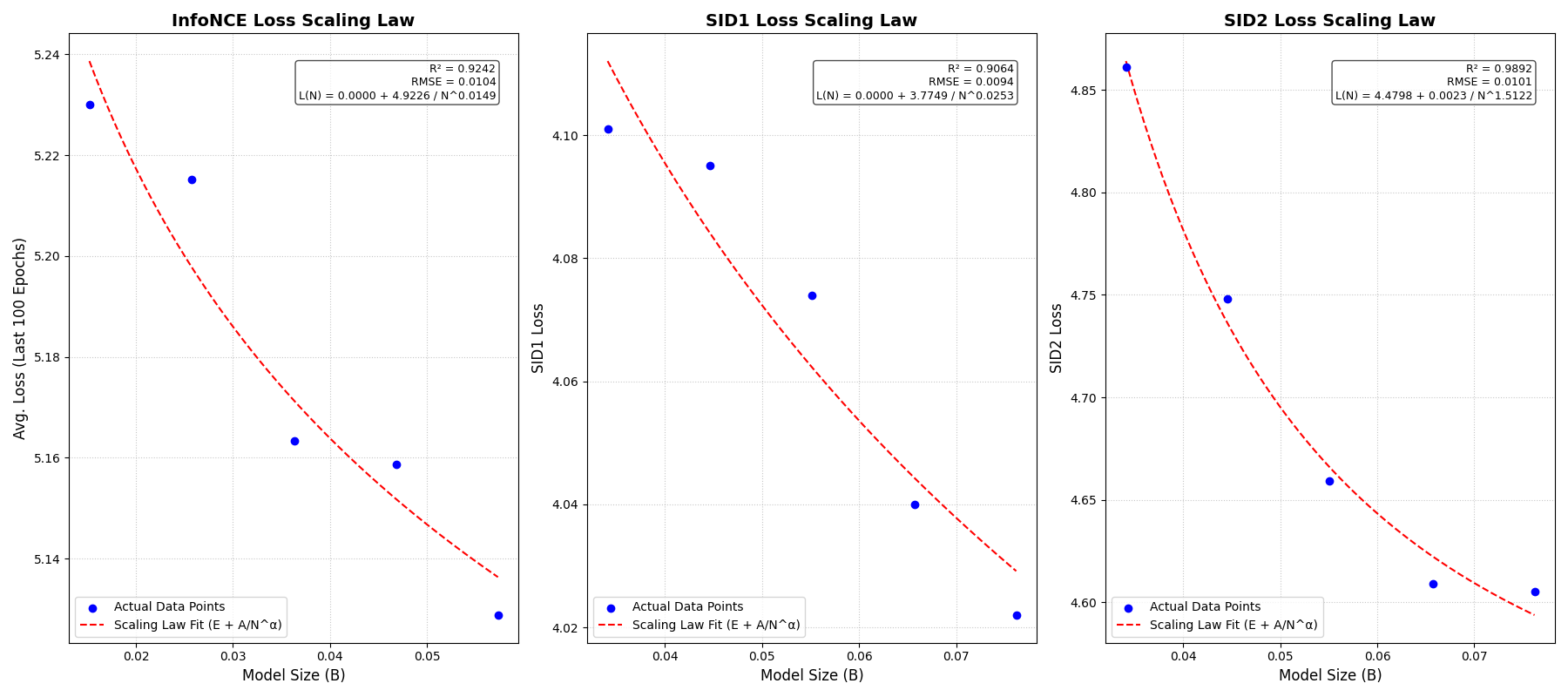}
\caption{The InfoNCE/Semantic ID prediction Scaling Laws we tested in TencentGR-100M dataset, first epoch.}
\label{scaling_law}
\end{center}
\end{figure}

\textbf{Our Work.}
To this end, we propose a unified encoder-decoder framework to connect them, while (1) the encoder has the SID1 and InforNCE losses; (2) the decoder has the SID2 loss under the SID1 condition.
Our design is grounded in a critical empirical observation (as illustrated in Figure~\ref{scaling_law}): within a unified backbone, the InfoNCE objective exhibits robust scaling behavior highly consistent with the generative objective (SID Loss), with both strictly adhering to power-law distributions ($R^2 > 0.9$).
Guided by this insight, we optimize both tasks within a shared \textbf{HSTU+MoE backbone} via a \textbf{coarse-to-fine strategy}:

\begin{enumerate}[leftmargin=*,align=left]
    \item \textbf{Stage 1: SID-Based Generative Retrieval.} 
    We use the backbone's generative capacity to rapid prune the search space via beam search (retrieving Top-$K$ SIDs). This step exploits the scaling laws of generative loss (SID1/SID2) to capture broad semantic structures.

    \item \textbf{Stage 2: Embedding-Based Scoring.} 
    We utilize the \textbf{same backbone} to output the user's embedding for precise dot-product scoring with LogQ debiasing technique. 
    Driven by our observation that embedding representation quality scales predictably with model size (InfoNCE scaling), this step effectively corrects generative hallucinations and ensures high-precision scoring as the model expands.
\end{enumerate}

\textbf{Contributions.}
The main contributions of our work are:

\begin{itemize}[leftmargin=*,align=left]
    \item \textbf{Unified Scaling Paradigm:} 
    We establish a unified architecture where semantic generation and embedding retrieval share a single scalable backbone, allowing both tasks to benefit simultaneously from Scaling Laws.
    \item \textbf{Collaborative Tokenizer:} 
    By fusing multi-modal and structural features, we design a residual quantization scheme that significantly reduces hash collisions (7.86\% for Top-50 items), providing high-resolution targets for the generative model.
    \item \textbf{Efficient Cascade Inference:} 
    The proposed cascade inference mechanism combines generative Retrieval with embedding refinement, addressing the ranking imprecision of pure generative models while maintaining efficient system response latency.
\end{itemize}

\section{Preliminary}
\subsection{Problem Statement}
Our task is to predict the next item $i_{t+1}$ a user $u \in \mathcal{U}$ is most likely to interact with, given their historical interaction sequence $S_u = (i_1, i_2, \dots, i_t)$, where $i \in \mathcal{I}$ is an item from the full corpus.
To enable scalable generative recommendation, we first define a semantic tokenizer $\mathcal{T}$ that maps each item $i$ to a set of hierarchical semantic codes (SIDs), $\mathbf{c}_i = (c_i^1, c_i^2)$.
Our cascade framework decomposes this problem into two sub-tasks:
\begin{itemize}
    \item \textbf{SID-Based Generative Retrieval:} 
    Given $S_u$, generate a high-quality candidate set of Top-$K$ SIDs.
    
    \item \textbf{Embedding-Based Scoring:} 
    Given $S_u$, compress its interests embedding $\mathbf{h}_{t+1}$, and use this embedding to perform precise scoring on the SID-retrieved candidate set.
\end{itemize}

\subsection{MLLM Embedding Analysis}
\begin{table}[htbp]
\centering
\caption{MLLM Embedding Coverage Statistics.}
\begin{tabular}{lccc}
\toprule
Member & Covered Items & Total Items & Coverage Rate \\
\midrule
81 & 16,693,655 & \multirow{5}{*}{19,099,627} & 87.403\% \\
82 & 16,732,879 & & 87.600\% \\
83 & 16,733,606 &  & 87.612\% \\
% 84 & 14375375 & 19099627 & 75.625\% \\
85 & 7,534,474 &  & 39.448\% \\
86 & 7,161,580 &  & 37.495\% \\
\bottomrule
\end{tabular}
\label{tab:coverage}
\end{table}
The raw multi-modal embeddings for item tokenization exhibit severe coverage inconsistencies. 
As shown in Table~\ref{tab:coverage}, some modalities (e.g., 82, 85) cover less than 40\% of the item corpus.
This sparsity renders any single modality unreliable for tokenization, as it would cause massive \textbf{out-of-vocabulary (OOV)} issues.
Addressing this data sparsity via a robust fusion strategy is therefore a critical prerequisite for building our generative model.

\section{Methodology}
 \begin{figure}[t!]
\begin{center}
\includegraphics[scale=0.3]{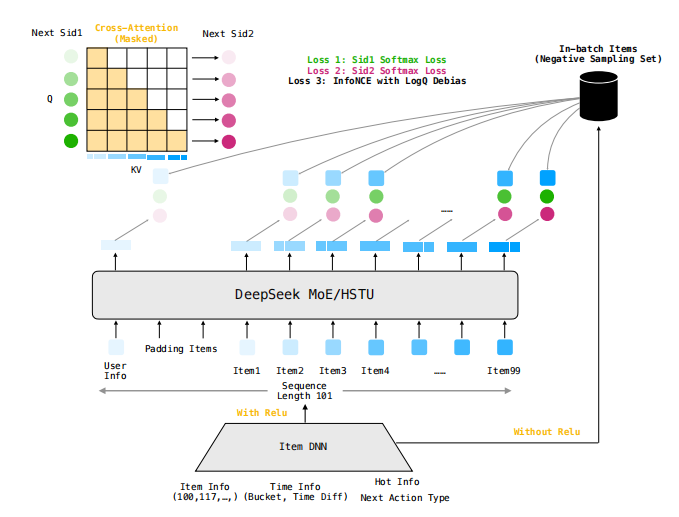}
\caption{The training workflow of OnePiece.}
\label{fig:ourmethod}
\end{center}
\end{figure}

The architecture of our framework (Figure~\ref{fig:ourmethod}) consists of three integral components: a \textbf{Semantic Tokenizer} for item discretization, a scalable \textbf{HSTU+MoE Sequence Encoder-Decoder}, and a \textbf{Hybrid Objective} mechanism.

\subsection{Semantic Tokenizer}
\label{sec:tokenizer}

% 表格
\begin{table*}[t]
\centering
\caption{\textbf{Collision Analysis of Semantic Tokenizer.} We compare single-modality embeddings against our \textbf{Collaborative} strategy, which is designed to resolve the low coverage and sparsity issues in individual modalities. 'Standard' denotes direct quantization, while 'Re-assigned' demonstrates the significant conflict reduction achieved by our \textbf{Greedy Re-assignment} strategy.}
\label{tab:res_kmeans_metrics}
\footnotesize
\resizebox{\textwidth}{!}{
\begin{tabular}{@{}lcccccccc@{}}
\toprule
\multirow{3}{*}{\textbf{Modality}} & 
\multicolumn{2}{c}{\textbf{Training Loss}} & 
\multicolumn{3}{c}{\textbf{Standard Quantization}} & 
\multicolumn{3}{c}{\textbf{w/ Greedy Re-assignment}} \\
\cmidrule(lr){2-3} \cmidrule(lr){4-6} \cmidrule(lr){7-9}
& \textbf{Loss1} & \textbf{Loss2} & \textbf{Conflicts} & \textbf{Conflict Rate} & \textbf{Unique Pairs} & \textbf{Conflicts} & \textbf{Conflict Rate} & \textbf{Unique Pairs} \\
\midrule
81 & 0.0184 & 0.0085 & 12.6M & 75.88\% & 4.02M & 2.15M & 12.88\% & 14.54M \\
82 & 0.2722 & 0.2220 & 13.7M & 81.95\% & 3.02M & 1.37M & 8.19\% & 15.36M \\
83 & 0.1653 & 0.1291 & 14.9M & 89.43\% & 1.76M & 7.11M & 42.51\% & 9.61M \\
85 & 0.1545 & 0.1177 & 5.45M & 72.34\% & 2.08M & 0.49M & 6.51\% & 7.04M \\
86 & 0.1608 & 0.1262 & 5.36M & 74.98\% & 1.79M & 0.59M & 8.34\% & 6.56M \\
\midrule
\textbf{Collaborative} & 0.4179 & 0.2968 & 14.4M & 75.46\% & 2.27M & 1.49M & \textbf{7.86\%} & \textbf{17.60M} \\
\bottomrule
\end{tabular}}
\end{table*}

We employ \textbf{Residual K-means} to quantize item $i$ into $\mathbf{c}_i = (c_i^1, c_i^2)$ with codebooks of size $16,384$. 
Since single-modality embeddings suffer from limited coverage (leading to high collisions), we propose a \textbf{Collaborative} strategy~\footnote{Utilizing an InfoNCE-only model to produce Item embedding.} to fuse multi-modal signals. 
This dense representation enables our \textbf{Greedy Re-assignment} mechanism (searching \textbf{Top-50} neighbors) to effectively resolve conflicts, achieving the highest resolution ($17.6$M unique pairs) and a minimal conflict rate of $7.86\%$, as detailed in Table~\ref{tab:res_kmeans_metrics}.

\subsection{Feature Engineer with ItemDNN}
The \textbf{ItemDNN} module fuses multifaceted item features (static info, time info, hot info) via an MLP.
As shown in Figure~\ref{fig:ourmethod}, we use a dual-path (ReLU and identity) design to preserve both non-linear ability for the Transformer input and identity-mapping raw feature signals for the faiss index building, to ensure the [-1, 1] cosine similarity space is fully used.

\subsection{Sequence Encoder}

\subsubsection{HSTU Backbone}
We adopt \textbf{HSTU}~\cite{zhai2024actions} as the backbone, which, unlike standard $O(L^2)$ Transformers, achieves near-linear scaling to efficiently process long user sequences.

\subsubsection{Sparse Mixture-of-Experts (MoE)}
To efficiently scale capacity, we also implement the \textbf{Sparse MoE} layers for a large parameters tuning with limited training resource (e.g., 7-card H20(96GB)). The layer dynamically routes input $\mathbf{x}$ to the top-$k$ experts:
\begin{equation}
    \text{MoE}(\mathbf{x}) = \sum_{j=1}^{N} g_j(\mathbf{x}) E_j(\mathbf{x}),
\end{equation}
where $g(\cdot)$ is the gating function.

\begin{figure}[th!]
\begin{center}
\includegraphics[width=8cm,height=2.7cm]{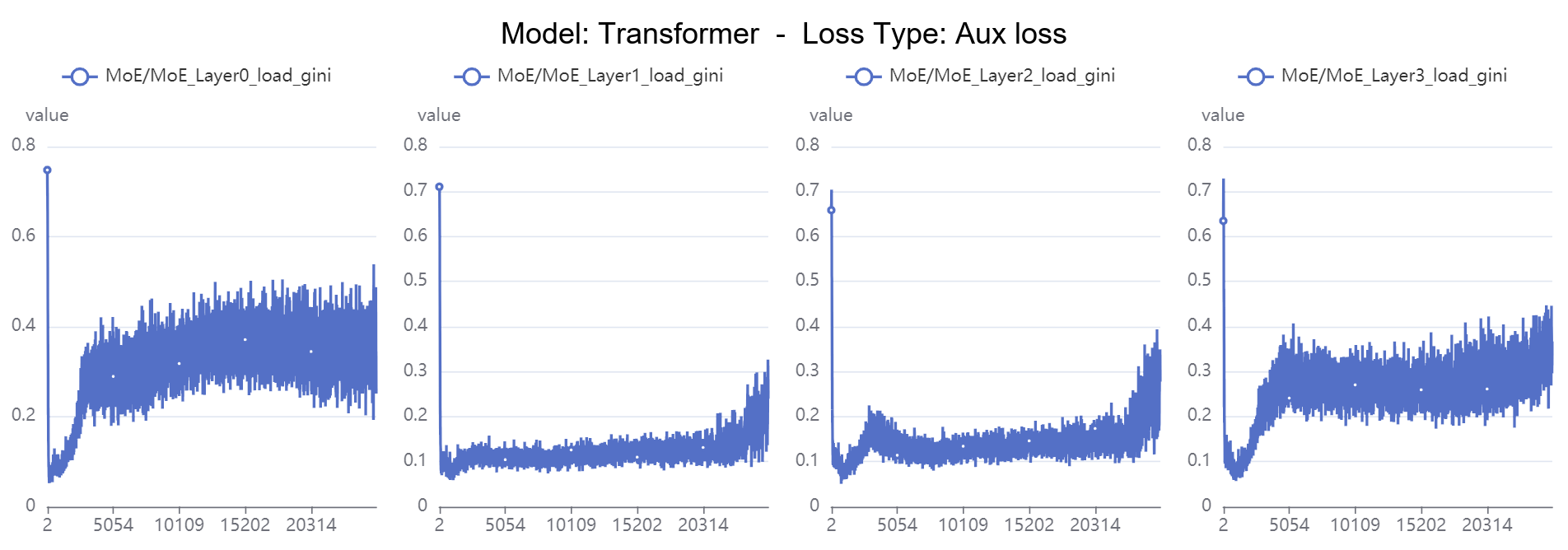}
\caption{Gini coefficient over training steps for all MoE layers. The rapid drop and stabilization at low values (0.1--0.4) indicate successful expert load balancing.}
\label{fig:gini}
\end{center}
\end{figure}

\textbf{Load Balancing.} To prevent expert collapse, we monitor the Gini coefficient of expert usage. As shown in Figure~\ref{fig:gini}, the rapid descent to a low stable range (0.1--0.4) across all layers confirms effective \textbf{load balancing}, fully leveraging the expanded capacity.

% Gini 图

\subsection{Hybrid Training Objectives}

Our framework optimizes a joint objective that unifies representation learning with generative semantic modeling.

\subsubsection{Embedding Matching with Debiasing Correction}
To align the sequence representation $\mathbf{h}_T$ with the target item $\mathbf{i}^+$, we employ the InfoNCE loss. 
Crucially, to mitigate the popularity bias inherent in in-batch sampling, we incorporate a \textbf{LogQ correction}~\cite{yi2019sampling}:
\begin{equation}
    \mathcal{L}_{con} = - \log \frac{\exp(\mathbf{h}_T^\top \mathbf{e}_{i^+} - \log Q(i^+))}{\sum_{j \in \mathcal{B}} \exp(\mathbf{h}_T^\top \mathbf{e}_{j} - \log Q(j))},
\end{equation}
where $Q(i)$ denotes the estimated item popularity and $\mathbf{e}$ represents item embeddings.

\subsubsection{Hierarchical Generative Modeling}
We introduce a generative task to predict the discrete codes $\mathbf{c} = (c^1, c^2)$ of the target item. This is formulated as a hierarchical auto-regressive process:
\begin{equation}
    p(\mathbf{c}|\mathbf{h}_T) = p(c^1|\mathbf{h}_T) \cdot p(c^2|c^1, \mathbf{h}_T).
\end{equation}
Specifically, the first-level code $c^1$ is predicted by attending to the encoder output $H$ using $\mathbf{h}_T$ as the query. 
For the second level, to enforce semantic consistency, we employ a \textbf{Teacher-Forcing} strategy: the query $\mathbf{q}_2$ is constructed by fusing $\mathbf{h}_T$ with the embedding of the first code, i.e., $\mathbf{q}_2 = \text{MLP}([\mathbf{h}_T; \mathbf{E}(c^1_{\text{gt}})])$, where $\mathbf{E}(c^1_{\text{gt}})$ denotes the embedding representation of the ground-truth code $c^1$ from the codebook.
Both probabilities are optimized via cross-entropy losses, denoted as $\mathcal{L}_{c^1}$ and $\mathcal{L}_{c^2}$.

\subsubsection{Joint Optimization}
The final objective balances the retrieval accuracy and semantic reconstruction:
\begin{equation}
    \mathcal{L}_{\text{total}} = \mathcal{L}_{con} + \lambda_1 \mathcal{L}_{c^1} + \lambda_2 \mathcal{L}_{c^2}.
\end{equation}

\section{Experiments}
\subsection{Inference Setting}
\begin{figure}[t!]
\begin{center}
\includegraphics[scale=0.25]{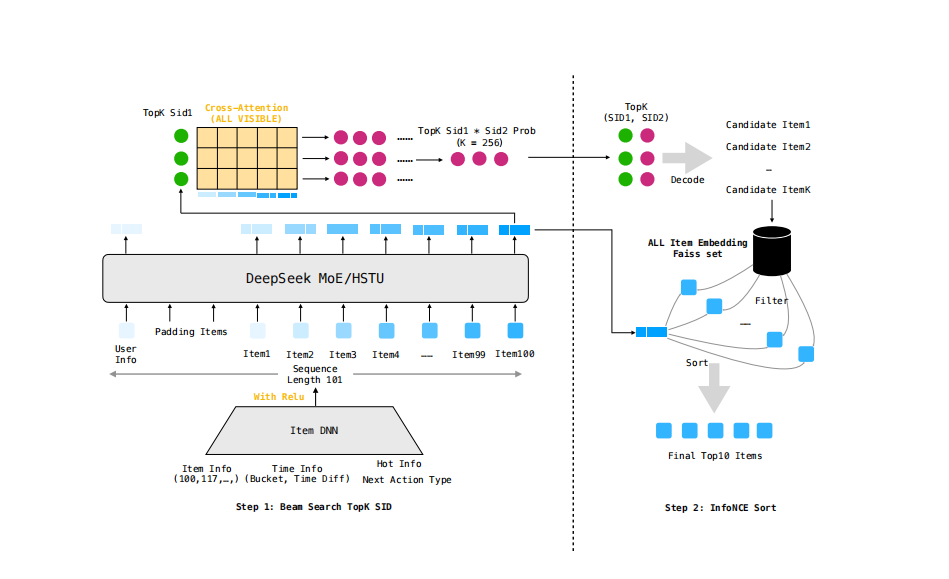}
\caption{The hybrid inference pipeline of OnePiece(SID beam search + InfoNCE Scoring).}
\label{fig:model_architecture}
\end{center}
\end{figure}

Our inference process, illustrated in Figure \ref{fig:model_architecture}, co-utilizes the model's auto-regressive (SID) and dual-tower (InfoNCE) capabilities.

\begin{enumerate}[leftmargin=*,align=left]
    \item % 1. 
    \textbf{Stage 1: Auto-Regressive Candidate Generation (SID Beam Search).}
    For a given user, we first use the model's auto-regressive head to decode the most probable $(\text{sid1}, \text{sid2})$ sequences via Beam Search. We set the beam width to $B=20$. We compute the joint log-probability $P(\text{sid1}, \text{sid2})$ and select the Top-K' SID pairs with the highest scores. Here, K' = 384. We then reverse-map these SIDs to retrieve a candidate items, denoted as $C_{\text{sid}}$.
    \item % 
    \textbf{Stage 2: Dual-Tower Scoring (InfoNCE Scoring).}
    We compute the user's query embedding $q_u$ and retrieve the item embeddings for all items within the candidate set $C_{\text{sid}}$ generated in Stage 1. Instead of using the beam search generation scores, we re-rank the items in $C_{\text{sid}}$ using the cosine similarity scores $s(q_u, k_i)$ from the InfoNCE dual-tower model.
    \item % 
    \textbf{Stage 3: Filtering Strategy.}
    Before returning the final Top-10 results, two key filtering strategies are applied:
    \begin{itemize}[leftmargin=*]
        \item \textbf{Historical Behavior Filtering:} We explicitly filter out any item that has already appeared in the user's historical interaction sequence by setting its re-ranking score to $-\infty$.
        \item \textbf{Cold-Start Item Filtering:} During the creation of the full candidate embedding matrix, all cold-start items (i.e., items not present in the training set) are discarded.
    \end{itemize}
\end{enumerate}

The final recommended list consists of the Top-10 items from $C_{\text{sid}}$ after InfoNCE re-ranking and filtering.

\subsection{InfoNCE with Different Layers}
To validate the scaling properties of our model, we evaluated the performance of the InfoNCE ranking task across different model depths.

We fixed other key hyperparameters: the model is an HSTU architecture with a hidden dimension $D=128$, 8 attention heads, and Pre-LayerNorm.The maximum sequence length was 101 with a per-device batch size of 512.We used pure bfloat16 for training.For optimization, we employed the Muon optimizer with a learning rate of 0.08, momentum of 0.95, and a cosine annealing schedule with 2000 warmup steps. The InfoNCE loss used cosine similarity and a fixed temperature $\tau=0.02$. We then varied the number of layers in the sequence encoder, ranging from 8 to 40.

As shown in Table \ref{tab:final_model_layer_metrics}, we recorded the average training metrics for the last 100 batches after 1 epoch. The results clearly indicate a stable improvement in ranking performance (Hitrate and NDCG) as the number of layers increases, confirming our scaling hypothesis.

\begin{table}[htbp]
\centering
\footnotesize
\caption{Final Model Performance with Different Layer Configurations (Avg. of Last 100 Batches, 1-Epoch)}
\label{tab:final_model_layer_metrics}
\begin{tabular}{lccc}
\toprule
\textbf{Layer Configuration} & \textbf{Model Size} & \textbf{Hitrate} & \textbf{NDCG} \\
\midrule
40 Layers & 0.0573B & 0.3219 & 0.1812 \\
32 Layers & 0.0468B & 0.3153 & 0.1752 \\
24 Layers & 0.0363B & 0.3027 & 0.1706 \\
16 Layers & 0.0257B & 0.2936 & 0.1672 \\
8 Layers & 0.0152B & 0.2770 & 0.1579 \\
\bottomrule
\end{tabular}
\end{table}

Furthermore, we trained our optimal [40]-layer configuration (0.0573B) for multiple epochs to observe its convergence and overfitting behavior. Table \ref{tab:combined_epoch_performance} details the validation/test metrics across epochs. While training metrics (like Hitrate) consistently rise, the validation metrics (Validation NDCG) peak at Epoch 6 (0.1064) before showing signs of slight overfitting.

\begin{table}[htbp]
\centering
\caption{Valid/Test Performance Comparison across Epochs.}
\label{tab:combined_epoch_performance}
\footnotesize
\setlength{\tabcolsep}{4pt} 
\resizebox{\linewidth}{!}{
    \begin{tabular}{l cc cc}
    \toprule
    \multirow{2}{*}{\textbf{Epoch}} & 
    \multicolumn{2}{c}{\textbf{Valid}} & 
    \multicolumn{2}{c}{\textbf{Test}} \\
    \cmidrule(lr){2-3} \cmidrule(lr){4-5}
    
    & \textbf{HR@10} & \textbf{NDCG@10} & \textbf{HR@10} & \textbf{NDCG@10} \\
    \midrule
    5 & 0.1885 & 0.1062 & 0.1307 & 0.0717 \\
    6 & \textbf{0.1888} & \textbf{0.1064} & 0.1313 & 0.0721 \\
    7 & 0.1885 & 0.1061 & \textbf{0.1316} & \textbf{0.0723} \\
    8 & 0.1879 & 0.1057 & 0.1313 & 0.0722 \\
    \bottomrule
    \end{tabular}
}
\end{table}
\subsection{SID with Different Layers}
We also evaluated the scaling properties of the auto-regressive SID generation task.
We fixed key hyperparameters: the model is an HSTU architecture, $D=128$, 8 attention heads, Pre-LN, AdamW optimizer with $\text{lr}=0.001$, cosine annealing with 2000 warmup steps, and the InfoNCE loss was disabled.

As shown in Table \ref{tab:layer_performance}, increasing model depth (from 4 to 20 layers) consistently improves the accuracy (HR@10) for SID1/SID2.

\begin{table}[htbp]
\centering
\footnotesize
\caption{HitRate@10 Performance with Different Layer Configurations (1 Epoch, Last 100 Batches Avg)}
\label{tab:layer_performance}
\begin{tabular}{lccc}
\toprule
\textbf{Layer Configuration} &\textbf{Model Size}&  \textbf{SID1 HR@10} & \textbf{SID2 HR@10} \\
\midrule
20 Layer &0.0762B& 0.5804 & 0.4419 \\
16 Layers &0.0657B& 0.5761 & 0.4345 \\
12 Layers & 0.0551B&0.5759 & 0.4280 \\
8 Layers & 0.0446B&0.5749 & 0.4149 \\
4 Layers & 0.0341B&0.5711 & 0.3979 \\
\bottomrule
\end{tabular}
\end{table}

We also trained the [20]-layer configuration for multiple epochs and applied our hybrid inference strategy (SID Beam Search + InfoNCE ranking) for evaluation. Table \ref{tab:combined_epoch_performance} shows the validation results. The performance peaks at Epoch 7 (NDCG@10 of 0.0723), suggesting the model achieves a balance of SID generation and InfoNCE ranking capabilities at this point.

\section{Conclusion}
This work demonstrates that the semantic generation and in-batch InfoNCE can be synergistically scaled within a unified backbone.
Through our cascade framework, we empirically validated that both tasks adhere to power-law Scaling Laws, and effectively resolved high-frequency item collisions via a collaborative tokenizer.
Future work will explore the behavior of Scaling Laws on larger-scale (e.g., billion-parameter) multi-modal foundation models and investigate end-to-end differentiable optimization strategies to further unify the training process.

\section*{Acknowledgement}
We thank the Tencent Advertising Algorithm competition (TAAC) organizers for providing several months the 7-card H20(96GB) computing resources and industrial-scale TencentGR-10M/100M datasets to help us explore the generative recommendation.

% \newpage
\balance
\bibliographystyle{ACM-Reference-Format}
\bibliography{sample-base-extend.bib}

\end{document}